\documentclass[pre,showpacs,twocolumn,superscriptaddress]{revtex4-1}
\usepackage{amssymb,graphicx,amsbsy,amsmath}

\begin{document}
\title{Nonequilibrium steady state of the kinetic Glauber-Ising
model under an alternating magnetic field}
\author{Seung Ki Baek}
\email[E-mail: ]{seungki@pknu.ac.kr}
\affiliation{Department of Physics, Pukyong National
University, Busan 608-737, Korea}
\affiliation{School of Physics, Korea Institute for Advanced Study,
Seoul 130-722, Korea}
\author{Fabio Marchesoni}
\affiliation{School of Physics, Korea Institute for Advanced Study,
Seoul 130-722, Korea}
\affiliation{Dipartimento di Fisica,
Universit{\`a} di Camerino, I-62032 Camerino, Italy}

\begin{abstract}
When periodically driven by an external magnetic field, a spin system
can enter a phase of steady entrained oscillations with
nonequilibrium probability distribution function. We consider an
arbitrary magnetic field switching its direction with frequency
comparable with the spin-flip rate and show that the resulting
nonequilibrium probability distribution can be related to the system
equilibrium distribution in the presence of a constant magnetic field
of the same magnitude. We derive convenient approximate expressions
for this exact relation and discuss their implications.
\end{abstract}


\maketitle

\section{Introduction}

The equilibrium properties of a statistical-physical system are often
characterized by a few macroscopic degrees of freedom. As the system
gets out of equilibrium, however, a huge, mostly unmanageable number
of degrees of freedom come into play. For this reason, most
conventional approaches to nonequilibrium physics have recourse to
the linear-response approximation, where the response of the system
to a small perturbation is expressed in terms of equilibrium
properties. The possibility of an exact formalism incorporating
nonequilibrium processes has recently emerged with the discovery of
the so-called fluctuation theorems~\cite{evans,*cohen,*jar,*crooks}
and the formulation of steady-state
thermodynamics~\cite{oono,*hatano,*koma,*koma2}. Popular study cases
of collective nonequilibrium dynamics are provided by classical spin
models, such as the kinetic Glauber-Ising model~\cite{glauber}.
In addition to the earlier literature, where the dynamical phase
transitions in such low dimensional stylized systems have been
investigated at
depth~\cite{sides2,*sides1,*sides3,korniss2,*korniss1,robb,park1,*park2}, we
focus here on a different aspect of the problem, namely on the search for an
algebraic
framework to characterize a nonequilibrium steady state (NESS).
This class of systems can be maintained out of equilibrium by a variety of
external agents, like multiple heat reservoirs~\cite{zia} or external
time-dependent magnetic fields~\cite{tome90}. For instance, when a
weak, slowly oscillating magnetic field is applied to the
Glauber-Ising model, the system eventually enters a steady collective
oscillation phase via entrainment. The linear-response theory
accurately describes the onset of entrainment by adopting the
average magnetization as an order parameter~\cite{leung,*double,*double2}.
However, we show below that
such a perturbation approach fails to determine the probability
density function (PDF) itself or other observables that are nonlinear
functions of the PDF, like the entropy.

The approach pursued in this work is opposite to the
linear-response theory: Instead of restricting ourselves to the
low-frequency regime, where the magnetic field oscillates with a period
much longer than the spin-flip time scale, here we assume from the
beginning a high-frequency regime, where the driving frequency and
the spin-flip rate are comparable. We show that, even if this
situation occurs far from equilibrium, there exists a rather simple
relationship between the NESS for the
driven spin system, and the known Boltzmann equilibrium PDF for the
system subject to a constant magnetic field. This result can be
then extended to analyze more realistic situations for lower driving
frequency. In this first report, we focus on globally coupled spin
systems, whose critical behavior belongs to the mean-field (MF)
universality class. In view of practical applications, we remind that
this is the universality class of three-dimensional quantum Ising
ferromagnets and uniaxial dipolar Ising ferromagnets~\cite{guggen1,*nielsen}.

This work is organized as follows: In Sec.~\ref{sec:pert}, we attempt a
perturbative approach to obtain the NESS under sinusoidal
modulation, and compare it with numerical results. In Sec.~\ref{sec:algebra},
we present an alternative algebraic formulation for square-wave
modulation at high frequency, yielding the NESS as an eigenvector.
We derive an approximate expression at lower frequencies as well. After
comparing our formula with numerical results, we summarize this work in
Sec.~\ref{sec:summary}.

\section{Perturbative approach}
\label{sec:pert}

Let us consider $n$ Ising spins governed by the Glauber dynamics. The
number of possible configurations is $N \equiv 2^n$. For each spin
configuration $i = (\sigma_1, \ldots, \sigma_n)$, the energy function
is
\begin{equation}
E_i = - J \sum_{\left< \mu \nu \right>} \sigma_\mu \sigma_\nu - h
\sum_\mu \sigma_\mu, \label{eq:energy}
\end{equation}
where the first summation runs over the nearest neighbors and $h$ is
an external magnetic field. In the globally coupled case discussed
here, every spin is coupled to all the other spins so that the first summation
should be understood as running over all the spin pairs. At the same time,
the coupling strength $J$ is replaced by $J_0(n-1)^{-1}$, with
$J_0$ a constant, to ensure that the energy is an extensive quantity.
According to the Glauber dynamics, the transition rate from the spin
configurations $i = (\sigma_1, \ldots, \sigma_\alpha, \ldots
\sigma_n)$ to $j = (\sigma_1, \ldots, -\sigma_\alpha, \ldots
\sigma_n)$ is
\begin{equation}
w_{ji} = \frac{1}{2n} \left[1-\sigma_\alpha \tanh \left(\beta
J\sum_{k\neq \alpha} \sigma_k + \beta h \right)\right],
\label{eq:rate}
\end{equation}
with $\beta \equiv (k_B T)^{-1}$ and $T$ denoting the temperature of the heat
bath in contact with this system. To
simplify notation, in the following we set $J_0=1$ and $k_B=1$. The
prefactor $n^{-1}$ in Eq.~(\ref{eq:rate}) indicates that only one
spin was flipped. In terms of these transition rates, one can write
the master equation
\begin{equation}
\Delta p_i(t) = \Delta t \sum_{j \neq i}^N [w_{ij}(h) p_j(t) -
w_{ji}(h) p_i(t)], \label{eq:kmaster}
\end{equation}
where $p_i$ is the probability to observe the configuration $i$ and
$\Delta t$ is the average spin-flip time. The system PDF, denoted by
the vector $\mathbf{p}$, with transpose $\mathbf{p}^T = (p_1, \ldots,
p_N)$, is normalized to $1$, i.e., $\sum_{j=1}^N p_j = 1$.
This is one of the simplest systems exhibiting nontrivial collective behavior
such as dynamic phase transitions and hysteresis~\cite{rmp}. If the external
field is absent, the phase transition occurs at $T=1$ in units of $J_0/k_B$
in the thermodynamic limit.

\begin{figure}
\includegraphics[width=0.48\textwidth]{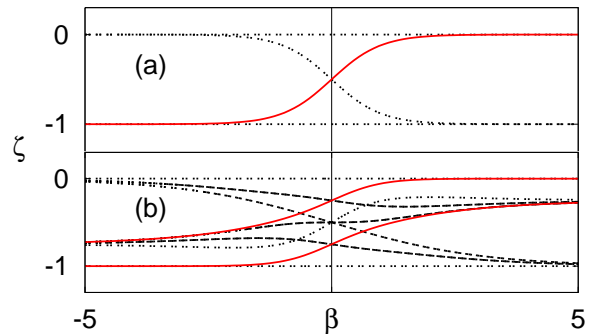}
\caption{(Color online) Eigenvalue spectrum of the transition matrix
at $h=0$, $\tilde{W}^\ast$,  for (a) $n=2$ and (b) $n=4$. The solid
lines represent the eigenvalues coupled to $h$, according to the
linear-response theory (see text). The negative-$\beta$ side
represents the antiferromagnetic Ising model~\cite{vives}.}
\label{fig:couple}
\end{figure}

We show first that standard linear perturbation analysis fails to
reproduce the $h$ dependence of $\mathbf p$, even for very small
system sizes. For a system of two spins, $n=2$, there exist $N= 4$
possible states, namely, $++, +-, -+$, and $--$. Equivalently, we
label these states $3,2,1$, and $0$, by digitizing the spin directions
$+$ and $-$, respectively, as $1$ and $0$. At low fields, $\beta h
\ll 1$, the transition rates $w_{ji}$ can be expanded in powers of
$\beta h$, so that $p_i(t)$ deviates from its equilibrium value,
$p_i^\ast$ at $h=0$, by a small amount $\eta_i$,
\begin{equation}
p_i(t) = p_i^\ast + \eta_i(t), \label{eq:perturb}
\end{equation}
with $p_3^\ast = p_0^\ast = [2(1+e^{-2\beta})]^{-1}$, $p_1^\ast =
p_2^\ast = [2(1+e^{2\beta})]^{-1}$, and $\sum_i \eta_i(t) = 0$. By
retaining all terms up to the first order in $\eta_i$ and $\beta h$,
the time evolution of $\mathbf{\eta}$, with $\mathbf{\eta}^T \equiv
(\eta_3, \eta_2, \eta_1, \eta_0)$, is governed by the linear equation
$d\mathbf{\eta}/dt = \tilde{W}^\ast \cdot \mathbf{\eta} +
\left(\frac{1}{4} \beta h~ {\rm sech}^2 \beta \right) \mathbf{\phi}$,
obtained by taking the limit $\Delta t \rightarrow 0$ in Eq.
(\ref{eq:kmaster}). Here, we have introduced the transition matrix at
$h=0$, $\tilde{W}^\ast$, and a coupling vector $\mathbf{\phi}$, with
$\mathbf{\phi}^T = (1, 0, 0, -1)$. The matrix $\tilde{W}^\ast$ has
eigenvalues $\zeta_3 = -1$, $\zeta_2 = 0$, $\zeta_1 =
\frac{1}{2}(-1-\tanh\beta)$, and $\zeta_0 =
\frac{1}{2}(-1+\tanh\beta)$, and the corresponding eigenvectors are
the columns of the diagonalization matrix $\tilde{Y}$.
After diagonalizing $W^\ast$ with $\tilde{Y}$, the equation
for $\eta_i(t)$ reads
\begin{equation}
\frac{d}{dt} \eta'_i = \zeta_i \eta'_i - \delta_{i0}\frac{\beta h}{4}{\rm
sech}^2 \beta,
\label{eq:couple}
\end{equation}
where the prime sign labels the transformed coordinates and
$\delta_{i0}$ is the Kronecker $\delta$ function. As $h(t)$ is
assumed next to vary slowly in time, in leading order, terms
proportional to $dh/dt$ can be safely discarded. In the case of
sinusoidally oscillating fields, $h(t) = h_0 \sin \omega t$, we can
easily solve the set of linear differential equations in Eq.
(\ref{eq:couple}) for large $t$ and transform the solutions back to
the original coordinates, namely, $\eta_1 = \eta_2 = 0$ and $\eta_0 =
\frac{1}{4}\beta h_0 {\rm sech}^2\beta (\omega \cos \omega t +
\zeta_0 \sin \omega t)/ (\zeta_0^2 + \omega^2) = -\eta_3$. Note that
$h(t)$ is only coupled to the eigenmode associated with the second
largest eigenvalue $\zeta_0$ [see Fig.~\ref{fig:couple}(a)]. At
larger $n$, the relaxation time toward $p_i^\ast$ is still determined
by the second largest eigenvalue $\zeta_0$ (i.e., the slowest
decaying mode) [Fig.~\ref{fig:couple}(b)]. As $n$ grows, the critical
point will roughly correspond to the resonance condition $|\zeta_0|
\approx \omega \rightarrow 0$, where the time scale diverges, so that
the ground state associated with $\zeta_2=0$ becomes doubly
degenerate.

\begin{figure}
\includegraphics[width=0.48\textwidth]{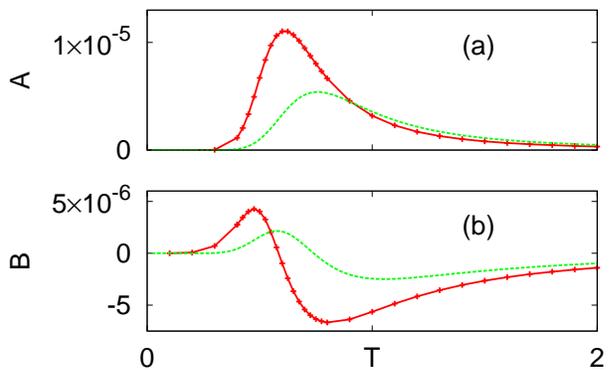}
\caption{(Color online) Amplitudes of entropy change for $n=2$,
$\omega=2\pi\times 10^{-2}$, and $h_0=10^{-2}$. The line points show
$A$ and $B$ of $d\langle S\rangle/dt$ [see
Eq. (\ref{eq:lin})] obtained by numerically
integrating Eq.~(\ref{eq:kmaster}) with $\Delta t=10^{-2}$, while the
dotted lines represent the corresponding analytic results
in the linear-response theory. } \label{fig:lin}
\end{figure}

We quantify now the system response to the external drive $h(t)$ by
calculating its entropy change as a function of
time~\cite{tome06,*tome10,*tome}. In this case, the nonequilibrium
entropy can be expressed as $\left< S \right> = -\sum_i p_i \ln p_i$
and approximated to $-\sum_i p_i^\ast \ln p_i^\ast -
\eta_3^2/p_3^\ast$. By inserting our estimate for $\eta_3$, we obtain
the rate of entropy change per spin,
\begin{equation}
\frac{1}{n} \frac{d\left< S \right>}{dt} \approx A \cos
2\omega t + B \sin 2\omega t, \label{eq:lin}
\end{equation}
where in the linear-response theory
$A^{\rm lin} \equiv -2\beta^2 e^{2\beta} \zeta_0 h_0^2 \omega^2
/ C$ and $B^{\rm lin} \equiv \beta^2 e^{2\beta} h_0^2 \omega
(-\zeta_0^2+\omega^2) / C$ with $C \equiv (1+e^{2\beta})^3
(\zeta_0^2+\omega^2)^2$. Since the system entropy is a periodic
function of time, differently from the entropy production of the
total process~\cite{tome06,*tome10,*tome}, the rate in
Eq.~(\ref{eq:lin}) has no definite sign. Note that, for a given
$\beta$, $A^{\rm lin}$ attains a maximum at $\omega = |\zeta_0|$, as
anticipated above. However, when compared with the numerical data
displayed in Fig.~\ref{fig:lin}, Eq.~(\ref{eq:lin}) clearly fails for
$\beta h \gtrsim O(10^{-2})$. The discrepancy gets even worse as the
system size increases. The failure of the linear-response theory is
consistent with the observation that at low $T$, in the large-$n$
limit, the system PDF may experience singular changes for
infinitesimal field modulations~\cite{goldenfeld}, which invalidates
the assumption of Eq.~(\ref{eq:perturb}) for $\beta h\ll 1$.

\section{Algebraic formulation}
\label{sec:algebra}

\subsection{High-frequency modulation}

We introduce now an alternative approach aimed at overcoming the limitations of
the linear-response theory.
The main idea is that the up-down symmetry will be generally broken in the
presence of the external field, even though the field is oscillating, so that it
is better to choose a symmetry-broken equilibrium state as our starting point to
study the NESS~\cite{*[{See }] [{, for a similar approach
to the Kondo effect in a NESS.}] jhong}. This can be best explained in terms of
linear algebra in the following way:
Let $U(h)$ denote the transition matrix for
a spin system of energy function as in Eq. (\ref{eq:energy}),
subject to an external magnetic field $h$. Under a static field $+h$,
the corresponding
system dynamics is formulated as an $N \times N$ matrix equation,
$\mathbf{p}(t+\Delta t) = U(+h) \cdot \mathbf{p}(t)$, with a
steady-state solution coinciding with the eigenvector associated with
the largest eigenvalue, $\lambda_1=1$, that is $\mathbf{q_1} = U(+h)
\cdot \mathbf{q_1}$. After normalization, this determines the system
equilibrium PDF at constant $h$. The existence and uniqueness of the
eigenvector $\mathbf{q}_1$ for any finite $n$ is ensured by the
Perron-Frobenius theorem~\cite{meyer}. We hereafter assume finite $n$
and full knowledge of the $U(+h)$ spectrum, i.e., of all eigenmodes
$\mathbf{q}_i$ as solutions of the matrix equation $U(+h) \cdot
\mathbf{q}_i = \lambda_i \mathbf{q}_i$, with $\mathbf{q}_i^T \cdot
\mathbf{q}_i = 1$ and $\lambda_i$ denoting the $i$th largest
eigenvalue.
If the field changes its sign at every time step,
$\Delta t$, {\it with constant magnitude}, then the time evolution of
the PDF obeys the equation
\begin{equation}
\mathbf{p}(t+2\Delta t) = U(-h) \cdot U(+h) \cdot \mathbf{p}(t).
\label{eq:2t}
\end{equation}
Equation~(\ref{eq:2t}) describes the fastest oscillating field that
a discrete-time formulation with time step $\Delta t$ can accommodate
(see, e.g., Ref.~\cite{work}). To make notation more compact, we
define $U^{\pm} \equiv U(\pm h)$. These two matrices are related by
a similarity transformation $U^- = P \cdot U^+ \cdot P$, where $P$
is a permutation matrix exchanging the $h$ direction from $+$ to $-$
and vice versa. Note that $P^2 = I$, $I$ being the identity matrix.
Accordingly, Eq.~(\ref{eq:2t}) can be rewritten as
$\mathbf{p}(t+2\Delta t) = P \cdot U^+ \cdot P \cdot U^+ \cdot
\mathbf{p}(t) = [P \cdot U^+]^2 \cdot \mathbf{p}(t)$. Under
steady-state conditions, the system PDF is given by the
solution $\tilde{\mathbf{p}}$ of the following equation:
\begin{equation}
\tilde{\mathbf{p}} = P \cdot U^+ \cdot \tilde{\mathbf{p}},
\label{eq:target}
\end{equation}
with the system alternating between $\tilde{\mathbf{p}}$ and
$P\cdot\tilde{\mathbf{p}}$ at every time step. When replacing $[P \cdot U^+]^2$
by $[P \cdot U^+]$ in the right-hand side (rhs)
of Eq.~(\ref{eq:target}), one might argue that
$\mathbf{p}(t+\Delta t) = \pm P \cdot U^+ \cdot \mathbf{p}(t)$; However as
all elements in $P$, $U^+$, and $\tilde{\mathbf{p}}$ are
non-negative, the $+$ sign is the correct choice.
Since $P$ is a known matrix and $U^+$ was assumed to be known, one
expects that the NESS, $\tilde{\mathbf{p}}$, and the equilibrium PDF
associated with $U^+$, $\mathbf{q_1}$, are algebraically related.
The desired relationship can be established by multiplying
Eq.~(\ref{eq:target}) times $P$ and subtracting $\tilde{\mathbf{p}}$ from
both sides to get $(U^+ - I) \cdot \tilde{\mathbf{p}} = (P-I) \cdot
\tilde{\mathbf{p}}$. Unfortunately, $(U^+ - I)$ is non-invertible
because the largest eigenvalue $\lambda_1=1$ requires $\det
(U^+-\lambda_1 I) = 0$. One circumvents this difficulty by analyzing
the subspace orthogonal to $\mathbf{q}_1$, i.e., rewriting
$\tilde{\mathbf{p}}$ as
\begin{equation}
\tilde{\mathbf{p}}=X_\epsilon \cdot \tilde{\mathbf{p}} + c~
\mathbf{q}_1, \label{eq:x}
\end{equation}
where the sparse matrix $\epsilon$ in the projection operator
$X_\epsilon \equiv (U^+ - I +\epsilon)^{-1} \cdot (P-I)$ is required
to make the inversion possible (see Drazin inverse in
Ref.~\cite{meyer}). The reason for the unknown $c$ in
Eq.~(\ref{eq:x}) is that this subspace retains no information about
the direction of $\mathbf{q}_1$. A convenient choice for $\epsilon$
is as follows. Let us define a block matrix $Q \equiv (\mathbf{q}_1,
\mathbf{q}_2, \cdots, \mathbf{q}_N)$ so that in the transformed coordinates,
$Q^{-1} \cdot (U^+ - I) \cdot Q$ is a diagonal matrix with the first diagonal
element $\lambda_1-1=0$.  The other diagonal elements are nonzero as long as
$\lambda_k < \lambda_1 = 1$ for $k>1$. To make the first diagonal element
nonzero, we then consider a matrix with a single nonzero element
$\epsilon'_{ij} = -\delta_{i1} \delta_{j1}$, which corresponds to
$\epsilon = Q \cdot \epsilon' \cdot Q^{-1}$ in the original coordinates. Now,
$Q^{-1}\cdot (U^+ - I + \epsilon) \cdot Q$ is clearly invertible, whereas
$Q^{-1}\cdot (U^+ - I) \cdot Q$ was not, so we have explicitly constructed
$X_\epsilon$.
It is important that $\lambda_1$ is no eigenvalue of
$X_\epsilon$, so that the solution of Eq.~(\ref{eq:x}),
\begin{equation}
\tilde{\mathbf{p}} = c~(I-X_\epsilon)^{-1} \cdot \mathbf{q}_1,
\label{eq:dt1}
\end{equation}
relating $\tilde{\mathbf{p}}$ to $\mathbf{q}_1$ is well defined.
Finally, the constant $c$ is determined by normalizing
$\tilde{\mathbf{p}}$; most remarkably one can show that $\mathbf{p}^{\rm eq}
\equiv c~ \mathbf{q}_1$ is also a normalized PDF. This shows how
$\tilde{\mathbf{p}}$ in nonequilibrium is related to the
equilibrium PDF.

Note that the vector $\tilde{\mathbf{p}}$ can be expressed as a
polynomial by multiplying it times the lowest common denominator of
all the $N$ elements and imposing the normalization condition only at
the final step; hence, $\tilde{\mathbf{p}} = \tilde{\mathbf{p}}^{(1)}
+ \tilde{\mathbf{p}}^{(2)} +\ldots+ \tilde{\mathbf{p}}^{(n)}$, with
$\tilde{\mathbf{p}}^{(k)} \propto (\Delta t)^k$.
The idea is to construct the NESS as a series solution with a small expansion
parameter $\Delta t$. Such summands are
related to one another,
\begin{eqnarray}
0 &=& (P-I) \cdot \tilde{\mathbf{p}}^{(1)}\label{eq:1},\\
(U^+-I) \cdot \tilde{\mathbf{p}}^{(1)} &=& (P-I) \cdot
\tilde{\mathbf{p}}^{(2)},\nonumber\\
&\vdots&\nonumber\\
(U^+-I) \cdot \tilde{\mathbf{p}}^{(n-1)} &=& (P-I) \cdot
\tilde{\mathbf{p}}^{(n)}\nonumber,\\
(U^+-I) \cdot \tilde{\mathbf{p}}^{(n)} &=& 0. \label{eq:n}
\end{eqnarray}
This set of equations can also be written as
\begin{equation}
(U^+-I) \cdot \tilde{\mathbf{p}}^{(k-1)} = (P-I) \cdot \tilde{\mathbf{p}}^{(k)},
\label{eq:i}
\end{equation}
with $\tilde{\mathbf{p}}^{(k)}\equiv 0$ if $k\le 0$ or $k>n$.
It is clearly seen that one obtains the original equation to solve
[Eq.~(\ref{eq:target})] when summing up both sides.
Since Eq.~(\ref{eq:n})
should have a solution proportional to $\mathbf{q}_1$, which is known to us by
assumption, one may attempt to proceed recursively from Eq.~(\ref{eq:n}) all
the way up to Eq.~(\ref{eq:1}). Still, the singular matrix $(U^+-I)$ does not
allow the direct inversion but leaves an undetermined component proportional to
$\mathbf{q}_1$ every time.
Adding up these recursive solutions with the undetermined parts, we
end up with our key result, Eq.~(\ref{eq:x}).
To avoid lengthy algebraic manipulations, we limit ourselves to a
hand-waving argument for the recursive Eq. (\ref{eq:i}). As the
matrix $U^+$ is of the form $U^+ = I + \Delta t~W$, with $W \equiv
\{w_{ij} \}$, multiplying $\tilde{\mathbf{p}}^{(k-1)} \propto (\Delta
t)^{k-1}$ by $(U^+ - I)$ raises the exponent of $\Delta t$ by $1$,
thus relating $\tilde{\mathbf{p}}^{(k-1)} \propto (\Delta t)^{k-1}$
to $\tilde{\mathbf{p}}^k \propto (\Delta t)^k$. In addition, the
matrix $(P-I)$ on the rhs guarantees that one recovers
Eq.~(\ref{eq:target}) when resumming both sides of Eq. (\ref{eq:i}).
The truncation of the recursive Eqs.~(\ref{eq:i}) at $k=n+1$ is a
consequence of the MF character of the model. Indeed, for
models with lower symmetry the number of recursive equations would be
larger than $n+1$. In particular, the last equation implies that
$\tilde{\mathbf{p}}^{(n)}$ is proportional to $\mathbf{p}^{\rm eq}$. In fact,
only $\tilde{\mathbf{p}}^{(k)}$ with $k \ge n$ can be made
proportional to $\mathbf{p}^{\rm eq}$ in a MF model with $n+1$
different energy levels: For Glauber's transition rates with $w_{ij}
\propto \exp[\beta(E_j - E_i)]$, it takes products involving $n$ such
factors to obtain a PDF proportional to $\exp(-\beta E_i)$.

\subsection{Lower-frequency modulation}
\begin{figure}
\includegraphics[width=0.48\textwidth]{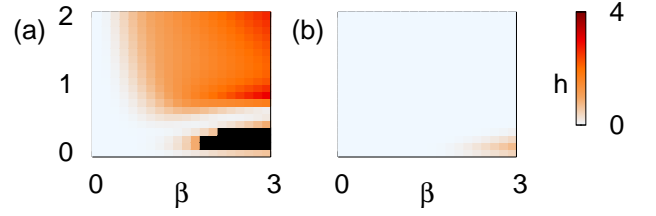}
\caption{(Color online) Kullback-Leibler divergence of
$\mathbf{p}^{\rm approx}$ from the exact PDF, $\tilde{\mathbf{p}}$
for $n=6$ spins; (a) $\gamma=10^1$ and (b) $\gamma=10^2$ with $\Delta
t \equiv 1$. The black
region in panel (a) denotes the parameter domain where
Eq.~(\ref{eq:approx}) breaks down (i.e., yields negative
probabilities).} \label{fig:comp}
\end{figure}

We extend now our analysis to lower driving frequencies by
considering the case when $h$ switches its sign every $\gamma$ time
steps, so that the NESS equation to solve is now
$\left(U^+\right)^\gamma \cdot \tilde{\mathbf{p}} = P \cdot
\tilde{\mathbf{p}}$. In the steady state, the system goes through the
transition sequences
\begin{eqnarray}
&\tilde{\mathbf{p}} \rightarrow U^+
\cdot \tilde{\mathbf{p}} \rightarrow \ldots \rightarrow (U^+)^{\gamma-1} \cdot
\tilde{\mathbf{p}} \rightarrow P \cdot \tilde{\mathbf{p}} \rightarrow
\nonumber\\
&U^- \cdot P \cdot \tilde{\mathbf{p}} \rightarrow \ldots \rightarrow
(U^-)^{\gamma-1} \cdot P \cdot \tilde{\mathbf{p}} \rightarrow
\tilde{\mathbf{p}} \rightarrow \ldots. \label{eq:seq}
\end{eqnarray}
As above, the steady-state solution is derived as $\tilde{\mathbf{p}} =
\left[ I-X_\epsilon(\gamma)\right ]^{-1} \cdot \mathbf{p}^{\rm eq}$, with
$X_\epsilon(\gamma) \equiv \left[(U^+)^\gamma-I+\epsilon \right]^{-1}
\cdot (P-I)$. Due to our choice for $\epsilon$ and using the Neumann
series $(A+B)^{-1} \approx A^{-1} - A^{-1}\cdot B\cdot
A^{-1}$~\cite{meyer}, we can approximate
$X_\epsilon(\gamma)$ to $X_\epsilon(\gamma) \approx \left[I +
(U^+)^\gamma + \epsilon\right] \cdot (I-P)$ and obtain
\begin{equation}
I-X_\epsilon(\gamma) \approx P + \left[ (U^+)^\gamma +\epsilon \right] \cdot
(P-I),
\label{eq:ix}
\end{equation}
as long as $\beta h \ll 1$ or $\gamma \gg 1$. In particular, on
increasing $\gamma$, the second term on the rhs of
Eq.~(\ref{eq:ix}) can be made much smaller than the first one. When
applied to $\mathbf{p}^{\rm eq}$, the inverse of the lhs of
Eq.~(\ref{eq:ix}) is then approximated, again through the Neumann
series, to
\begin{equation}
\tilde{\mathbf{p}} \approx P \cdot \mathbf{p}^{\rm eq} + P \cdot \left[
(U^+)^\gamma +\epsilon \right] \cdot(P-I)\cdot \mathbf{p}^{\rm eq}.
\label{eq:ixi}
\end{equation}
Therefore, the leading order of $[I-X_\epsilon(\gamma)]^{-1}$ is $P$,
and not $I$, even in the limit $h \rightarrow 0$, because $X_\epsilon
(\gamma)$ is not small compared to $I$~\cite{meyer}. The PDF
$\tilde{\mathbf{p}}$ should indeed be close to $P\cdot \mathbf{p}^{\rm eq}$
because $U^-$ has evolved the system for $\gamma$ time steps, so that
it is the second term on the rhs of Eq.~(\ref{eq:ixi}) that
describes the PDF change right after field reversal. Since $\left[
(U^+)^\gamma +\epsilon \right] \cdot \mathbf{p}^{\rm eq} = 0$, the dominant
change is proportional to $P \cdot \mathbf{q}_2$, whose elements add
up to zero. This is consistent with the predictions
(Fig.~\ref{fig:couple}) of the linear-response theory, which is
unable to distinguish between $\mathbf{q}_2$ and $P \cdot
\mathbf{q}_2$. We note that for $\beta \ll 1$ the matrix
$U^+$ is almost symmetric, which implies $\mathbf{q}_2^T \cdot
\mathbf{p}^{\rm eq} \ll 1$. Under these conditions a simple two-eigenmode
approximation allows us to go beyond the linear-response approximation,
by writing
\begin{equation}
\tilde{\mathbf{p}} \approx \mathbf{p}^{\rm approx} \equiv
P \cdot \mathbf{p}^{\rm eq} + \lambda_2^\gamma~
[\mathbf{q}_2^T \cdot (P-I) \cdot \mathbf{p}^{\rm eq}] (P\cdot \mathbf{q}_2).
\label{eq:approx}
\end{equation}
We checked the validity of this scheme by computing the
Kullback-Leibler divergence $D_{\rm KL}(\tilde{\mathbf{p}} ||
\mathbf{p}^{\rm approx}) \equiv \sum_{i=1}^N \tilde{p}_i \ln
\left(\tilde{p}_i/p^{\rm approx}_i
\right)$.  As displayed in Fig.~\ref{fig:comp}, with increasing
$\gamma$, $D_{\rm KL}$  decreases
over the whole parameter region. This confirms that in most cases the
perturbative description of Eq.~(\ref{eq:approx}) based on
the first two eigenmodes provides a reasonable approximation for
$\tilde{\mathbf{p}}$.

\begin{figure}
\includegraphics[width=0.48\textwidth]{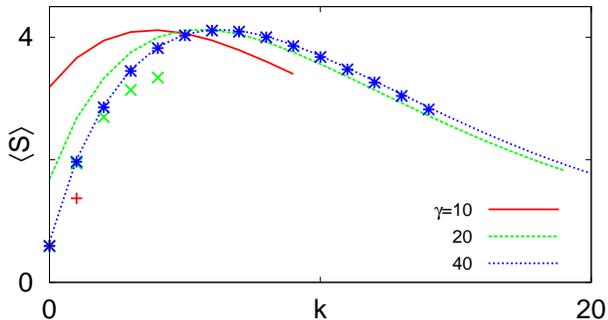}
\caption{(Color online) Nonequilibrium entropy as a function of the
number, $k$, of time steps with $\Delta t \equiv 1$,
for $n=6$ and $h = 1$ at temperature $T=1$. The solid
curves represent the exact results for different $\gamma$; the
crosses are approximate results
obtained by combining Eqs.~(\ref{eq:commutator}) and
(\ref{eq:evolution}) (see text), and the colors indicate the corresponding
values of $\gamma$.
Missing crosses at large $k$
indicate a breakdown of the approximation (negative predicted
probability).} \label{fig:commutator}
\end{figure}

Furthermore, when applied to $\mathbf{p}^{\rm eq}$, the commutator
$[U^+,P]$ can be estimated in terms of the first two PDF's,
$\tilde{\mathbf{p}}$ and $U^+\tilde{\mathbf{p}}$, in the transition
sequence of Eq.~(\ref{eq:seq}), i.e.,
\begin{equation}
[U^+,P] \cdot \mathbf{p}^{\rm eq}
\approx (U^+ - I) \cdot \tilde{\mathbf{p}}.
\label{eq:commutator}
\end{equation}
The time evolution of $\mathbf{p}^{\rm eq}$ is then formally expressed as
\begin{equation}
(U^+)^k \cdot P \cdot \mathbf{p}^{\rm eq} = \left\{ P + \left[ \sum_{j=0}^{k-1}
(U^+)^j \right] \cdot [U^+,P] \right\} \cdot \mathbf{p}^{\rm eq},
\label{eq:evolution}
\end{equation}
where we recall that for $\gamma \gg 1$ the lhs may be
approximated to $(U^+)^k \cdot \tilde{\mathbf{p}}$.
By using Eqs.~(\ref{eq:commutator}) and (\ref{eq:evolution}),
we numerically computed the time dependence of $\left< S \right>$
as plotted in Fig.~\ref{fig:commutator}, where this
approximation closely reproduces the numerical data at large
$\gamma$.

The $\Delta t$ power counting rule in Eq.~(\ref{eq:i}) can also be
generalized by considering
$\tilde{\mathbf{p}} = \tilde{\mathbf{p}}^{(1)} + \tilde{\mathbf{p}}^{(2)}
+\ldots+ \tilde{\mathbf{p}}^{(n\gamma)}$.
The matching condition for the orders of $\Delta t$ suggests
that Eq.~(\ref{eq:i}) be generalized to
\begin{equation*}
\sum_{k=0}^{\gamma}
\binom{\gamma}{k} (U^+-I)^k \cdot \tilde{\mathbf{p}}^{(i-k)} = P \cdot
\tilde{\mathbf{p}}^{(i)},
\label{eq:p}
\end{equation*}
where the binomial coefficients originate from combinatorial possibilities
in matching the orders.
The constraint is now given as $\tilde{\mathbf{p}}^{(i)}=0$ for $i\le 0$ or
$i>n\gamma$ in the MF case.
We note that the last $\gamma$ terms in the expansion are involved
only with $(U^+-I)$ so that they are always proportional
to the equilibrium solution.
We checked that the
symmetric part of $\tilde{\mathbf{p}}^{(1)}$ is independent of
$\gamma$ for small $n$, and this could be generic
because the $\tilde{\mathbf{p}}^{(1)}$ symmetry under $P$ [see
Eq.~(\ref{eq:i}) for $k=1$] implies its insensitivity to the field
direction. Therefore, the shapes of both the lowest-order,
$\tilde{\mathbf{p}}^{(1)} \propto \Delta t$, and the highest-order contributions
are independent of the external time scale $\gamma$.
If $\gamma$ is kept fixed, $\tilde{\mathbf{p}}$
becomes more symmetric with lowering $\Delta t$; accordingly, the
corresponding PDF turns out to be insensitive to $\gamma$ for
$\gamma \Delta t \ll 1$.

\section{Summary}
\label{sec:summary}

In summary, we have established an algebraic relationship between the NESS under
square-wave modulation and the equilibrium PDF under a constant magnetic field
of the same magnitude. Understanding a NESS is one of the most important
questions in nonequilibrium statistical physics, just as the Boltzmann
distribution forms the fundamental basis of the equilibrium statistical
mechanics.
It is particularly important in the specific context of the Glauber-Ising
model as well, because all the phenomena involved with the spontaneous symmetry
breaking in the dynamic phase transition at high frequency should be traced to
properties of the NESS.

We emphasize that the approach proposed here is not restricted solely
to the Glauber dynamics, but applicable to a general Markovian
system whose stationary state in the presence of a constant
external parameter is known; as the external parameter is
periodically modulated in time (with reflection symmetry), our
technique indicates how to express the NESS
in terms of the biased stationary state. An intriguing question is how
to extend our formalism to the case of a continuously varying field,
which requires approximating $h(t)$ to a piecewise constant function and
decoupling the eigenmodes at different times.

\acknowledgments We thank KIAS Center for Advanced Computation for
providing computing resources. This work was supported by the
Supercomputing Center/Korea Institute of Science and Technology
Information under Project No. KSC-2013-C1-004, and by the European
Commission under Project No. 256959 (NanoPower).

%
\end{document}